\documentclass[reprint,amsmath,amssymb,aps,prl,superscriptaddress,nofootinbib,floatfix]{revtex4-2}

\usepackage{graphicx}
\usepackage{bm}
\usepackage{mathrsfs}
\usepackage{amsmath}
\usepackage{latexsym}
\usepackage{array}
\usepackage{multirow}
\usepackage{booktabs}
\usepackage{tikz}
\usetikzlibrary{patterns,decorations.pathmorphing,shapes}
\usepackage[colorlinks=true, linkcolor=blue, urlcolor=blue, citecolor=magenta]{hyperref}
\graphicspath{{./}{figures/}{../figures/}}

\newcommand{\AO}{\mathrm{AO}}
\newcommand{\CS}{\mathrm{CS}}
\newcommand{\dd}{\mathrm{d}}
\newcommand{\ee}{\mathrm{e}}

\newcommand{\E}{\mathbb E}
\newcommand{\Prb}{\mathbb P}

\begin{document}

\title{Before the First Bond: Depletion-Assisted Stochastic Capture at the Onset of Endocytosis}

\author{Hao Wu}
\email{wuhao@ucas.ac.cn}
\affiliation{Zhejiang Key Laboratory of Soft Matter Biomedical Materials, Wenzhou Institute,
University of Chinese Academy of Sciences, Wenzhou, Zhejiang 325000, China}

\date{\today}

\begin{abstract}
The first barrier to receptor-mediated endocytosis is not binding but arrival within a few nanometers of the membrane. In a crowded extracellular environment, this pre-contact approach is driven by thermal fluctuations and entropic depletion, yet continuum models usually start from established contact or treat the approach deterministically. We formulate a thermodynamically consistent stochastic theory for this stage. The sphere--plane depletion geometry uses the exact Asakura--Oosawa spherical-cap overlap, not a fitted cap, with either ideal van~'t~Hoff or pressure-only Carnahan--Starling (CS) pressure. Near-wall mobility enters with its essential and often omitted noise-induced drift. The first specific reaction is a Robin radiation boundary; the outer boundary is reflecting or absorbing. Three first-passage quantities are obtained analytically by quadrature: the mean reaction time, the reaction-before-escape probability, and the conditional success time, the last encoded by a Doob transform that exposes the selection bias of conditioning on capture. The local Damk\"ohler number diagnoses the boundary reaction-to-diffusion ratio, while the mean-time crossover follows from the exact reaction and diffusion terms. At baseline, ideal and CS contact depths are \(2.33\) and \(5.95\,k_{\rm B}T\), and the reflecting-outer mean reaction time is approximately \(0.20\,{\rm ms}\); with an absorbing outer boundary the conditional success time is shorter, but the two observables use different boundary protocols. Numerical checks reproduce the analytic results and the radiation-boundary mass balance. This provides a quantitative theoretical framework for crowding-assisted capture before membrane wrapping.
\end{abstract}

\maketitle

\section{Introduction}

Endocytosis is the generic process by which eukaryotic cells internalize extracellular material, including nutrients, viruses, and engineered nanoparticles \cite{Doherty2009,zhang2015physical,wu2026unveiling}. In receptor-mediated endocytosis, the particle first approaches the membrane, then ligand-receptor bonds form, and finally the membrane wraps the particle \cite{Gao2005,Shi2008,mercer2010virus,Marsh2006}. Continuum models often begin from an already established contact or from a prescribed wrapping coordinate; the pre-contact approach is frequently compressed into a boundary condition or treated deterministically. At the nanoscale, thermal fluctuations, depletion forces, and finite surface reaction rates can occur on comparable time scales. This motivates an explicit stochastic treatment of the pre-contact stage of membrane wrapping.

Recent experimental and theoretical studies indicate that endocytosis can require an initial affinity \cite{Verma2010}. Such affinity may arise from entropic depletion forces generated by smaller molecules, including globular proteins and polymer coils \cite{Nelson2004,Asakura1954,Asakura1958,Vrij1976}. Depletion effects are well established in colloid science. Larger particles attract when the excluded volumes of smaller particles overlap, which increases the accessible volume of the smaller particles \cite{Dinsmore1995,Imhof1995,Steiner1995,Ilett1995}. Observations in vesicle systems \cite{Dinsmore1998} and in mammalian cells \cite{Irajizad2017} have been discussed in relation to depletion-driven membrane association. Reviews have emphasized entropy at the nano-bio interface \cite{Wan2024}. In crowded biological environments, crowding forces can drive macromolecular association without direct attraction \cite{Batra2009,Gupta2017,Dey2022,ellis2001crowding}. Entropic effects of membrane-anchored polymers can also modulate membrane mechanics and domain organization \cite{wu2013mechanical,wu2013effects}. Depletion forces are nevertheless relatively weak compared with specific ligand-receptor binding energies. Their primary role in endocytosis may be to reduce the effective barrier to initial contact rather than to drive wrapping by themselves.

A central geometric issue in applying depletion theory to endocytosis is the treatment of the depletion volume for finite-sized spheres near a planar membrane. Many treatments fit a spherical cap to the particle-membrane overlap. Here we instead use the exact cap obtained from the excluded-volume construction for crowder centers: the particle exclusion sphere of radius \(R+r\) overlaps the membrane-excluded half-space \(z<r\). This cap is exact for the ideal hard-sphere Asakura--Oosawa (AO) model before contact \cite{Miyazaki2022,Lekkerkerker2024}. Its analytic continuation into the wrapping phase is a separate approximation and is not exact once the membrane deforms. The osmotic pressure is treated either as the ideal van~'t~Hoff pressure or as a Carnahan--Starling (CS) pressure-only approximation \cite{CS1969}. Near-wall hydrodynamics and hindered diffusion also enter the normal gap coordinate, and the interpolated near-wall mobility used below is motivated by the Brenner and Bevan--Prieve results \cite{Brenner1961,BevanPrieve2000}.

\begin{figure}[t]
  \centering
  \includegraphics[width=\linewidth]{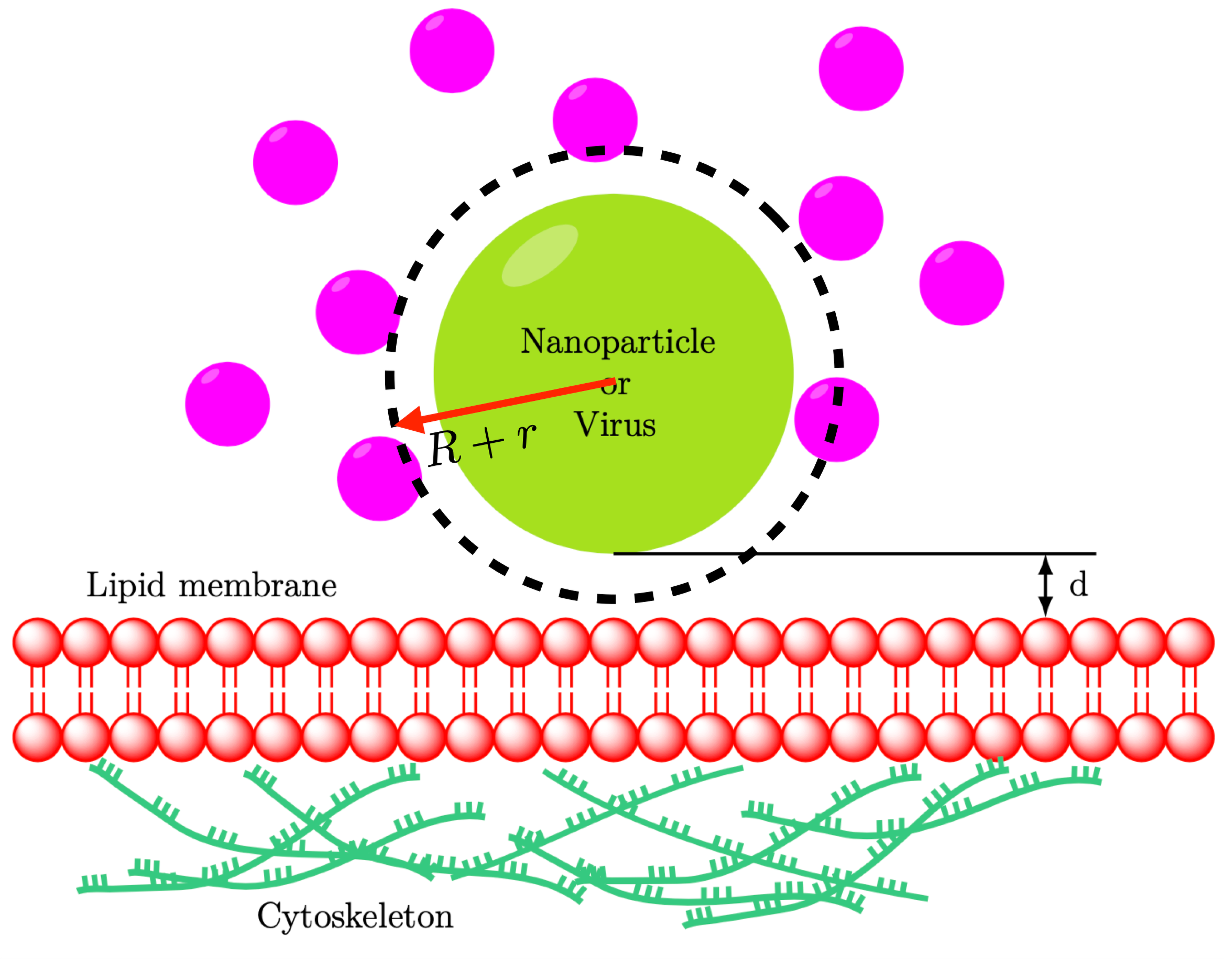}
  \caption{Cross-section of the particle, the crowder-center exclusion sphere of radius \(R+r\), the wall-excluded half-space \(z<r\), and the exact spherical-cap overlap. The cross-section marks \(z=0\), \(z=r\), \(z_C=R+d\), the radii \(R\) and \(R+r\), and the cap height \(q=2r-d\); the cap is obtained from the excluded-volume construction, not from a fitted spherical cap.}
  \label{fig:schematic}
\end{figure}

The cell is not purely elastic. Beneath the membrane, the cytoskeleton, composed of actin, microtubules, and intermediate filaments in a viscous cytosol, confers viscoelasticity with creep and stress relaxation on time scales from seconds to minutes \cite{mofrad2009cytoskeletal,Rotsch1999}. Treating the cell as a linear elastic half-space \cite{Sun2006} or ignoring cytoskeletal deformation can therefore misrepresent the energy landscape. Experimental disruption of actin dynamics alters viral entry rates \cite{matarrese2005human}. The observed optimal uptake size around \(50\,\mathrm{nm}\) in diameter \cite{Chithrani2006} has motivated models that include kinetic and non-equilibrium factors; this experimental diameter is not the model radius \(R\) used below. Recent experiments have measured forces during cellular uptake of viruses and nanoparticles at the ventral side \cite{WiegandEtAl2020}, and viscoelastic properties of basal plasma membranes and cortices have been characterized \cite{Janshoff2021}. Previous theoretical studies have incorporated cytoskeleton deformation to model engulfment kinetics after contact is established \cite{Zhang2021}, but they assume pre-existing contact and do not address initiation. Recent HIV modeling similarly assumes prior contact \cite{Kruse2023mathematical}. In this work, we examine entropic forces generated by nanoscale biomolecules as one possible contribution to the initiation step before specific binding. Once the particle is close, ligand-receptor binding sustains wrapping, and cytoskeletal viscoelasticity regulates kinetics through the Onsager variational principle \cite{Liu2026e1,Liu2026e2}. This principle minimizes the sum of the free-energy change rate and the Rayleigh dissipation function \cite{Onsager1931a,Onsager1931b,Doi2013}. It has been applied to polymer dynamics, colloids, and membranes \cite{Arroyo2009}. The elastic-viscoelastic correspondence principle extends Hertzian contact to viscoelastic media and incorporates time-dependent creep \cite{Lee1960,Radok1957}. Membrane bending and tension are described by the Helfrich--Canham Hamiltonian \cite{Helfrich1973,Canham1970,Ouyang1989,wu2018,wu2025generalized}, and the contact mechanics of the cytoskeletal response is treated within the standard linear solid framework \cite{Johnson1985}.

\begin{table*}[t]
\caption{Parameters used in the model and contextual parameters for the companion wrapping analysis. Only \(R\), \(r\), \(c\), \(T\), \(\mu\), and \(k_B\) enter the single-capture calculations of this paper. The receptor-ligand complex size \(\delta\) is recorded but is not identified with the capture distance \(d_c\). The adhesion energy density \(a\), bending modulus \(\kappa\), surface tension \(\gamma\), cytoskeletal viscosity \(\eta_c\), Young's moduli \(E_v\) and \(E_c\), and creep time \(\tau=\eta_c/E_c\) are listed for context and are not used in the present single-capture results. References or provenance are provided for each parameter.}
\label{tab:parameters}
\begin{tabular*}{\textwidth}{@{\extracolsep{\fill}}l l l@{}}
\toprule
\textbf{Parameter} & \textbf{Value} & \textbf{References} \\
\midrule
Virus radius \(R\) & \(50\)--\(100\) nm & \cite{Chithrani2006,Burrell2016,Liu2026e2} \\
Crowder radius \(r\) & \(5\)--\(20\) nm & \cite{Batra2009,Gupta2017,Kuznetsova2014,Dey2022,ellis2001crowding,Liu2026e2} \\
Crowder concentration \(c\) & \(1.5\times10^{22}\) m\(^{-3}\) & \cite{Mitchison2019,Alfano2024,Liu2026e2} \\
Receptor-ligand complex size \(\delta\) & \(5\) nm & \cite{Gao2005,Liu2026e2} \\
Total adhesion energy density \(a\) & \(2.9\)--\(5.8\times10^{-4}\) J\,m\(^{-2}\) & \cite{Sun2006,Liu2026e2} \\
Boltzmann constant \(k_B\) & \(1.380649\times10^{-23}\) J\,K\(^{-1}\) & Fundamental constant \\
Model temperature \(T\) & \(300\) K & Model temperature \\
Membrane bending modulus \(\kappa\) & \(8.18\times10^{-20}\) J & \cite{Li2012,Liu2026e2} \\
Membrane surface tension \(\gamma\) & \(2.07\times10^{-5}\) N\,m\(^{-1}\) & \cite{Sun2006,Li2024,Dai1998,Liu2026e2} \\
Water viscosity \(\mu\) & \(10^{-3}\) N\,s\,m\(^{-2}\) & Standard value \\
Cytoskeleton viscosity \(\eta_c\) & \(2696\) N\,s\,m\(^{-2}\) & \cite{Micoulet2005,Liu2026e2} \\
Virus Young's modulus \(E_v\) & \(10^8\) N\,m\(^{-2}\) & \cite{Sun2006,Li2012,Liu2026e2} \\
Cell Young's modulus \(E_c\) & \(2\times10^3\) N\,m\(^{-2}\) & \cite{Micoulet2005,Liu2026e2} \\
Creep retardation time \(\tau=\eta_c/E_c\) & \(1.348\) s & \cite{Liu2026e2}\\
\bottomrule
\end{tabular*}
\end{table*}

This paper is the first part of a two-paper analysis. The present part isolates the single-capture layer. A spherical particle approaches a planar membrane in a crowded extracellular environment. Ideal hard-sphere crowders generate an AO depletion attraction. The particle diffuses in the normal gap coordinate. A Robin boundary models the first specific reaction. The outer boundary is either reflecting or absorbing. The companion paper treats the subsequent multivalent bond stabilization after the first bond has formed. Related stochastic descriptions of receptor-mediated internalization have been developed in \cite{Billah2023,FedotovAlexandrov2024}. The first-passage and Robin-boundary machinery used here follows the standard stochastic-process and partially reflected diffusion literature \cite{Gardiner2004,Risken1989,Redner2001,LauLubensky2007,Singer2008,Grebenkov2023,GrebenkovSkvortsov2021}. Binding-competent residence times provide an order-of-magnitude reference \cite{hu2013binding}.

The physical parameters used in the single-capture calculations are collected in Table~\ref{tab:parameters}: \(R=50\)--\(100\,\mathrm{nm}\), \(r=5\)--\(20\,\mathrm{nm}\), \(c=1.5\times10^{22}\,\mathrm{m^{-3}}\), \(T=300\,\mathrm{K}\), and \(\mu=10^{-3}\,\mathrm{Pa\,s}\). The remaining entries in Table~\ref{tab:parameters} are contextual parameters for the companion wrapping analysis and are not used in the present single-capture results. The receptor-ligand complex size \(\delta=5\,\mathrm{nm}\) is recorded but is not automatically identified with the capture distance \(d_c\). The surface reaction speed \(k_s\), the capture distance \(d_c\), the initial position \(d_0\), and the initial distribution \(\rho_0(d)\) are conditional parameters. They are not Table~\ref{tab:parameters} inputs. All numerical values in this paper are therefore read together with the conditional-parameter table.

The exact AO sphere-plane excluded-volume geometry plays a central role in the theory. Its derivation is summarized in Appendix~\hyperref[app:AO]{A}. The membrane is the plane \(z=0\), with the crowder-accessible side \(z>0\). A particle of radius \(R\) has its lowest point at height \(d\ge0\). The particle-induced exclusion region for crowder centers is a sphere of radius \(A=R+r\) centered at \(z_C=R+d\). The wall-induced exclusion region is the half-space \(z<r\). For \(d<2r\), the AO overlap volume is the spherical cap
\begin{equation}
V_{\AO}(d)=\frac{\pi}{3}(2r-d)^2(3R+r+d),
\qquad 0\le d\le2r,
\label{eq:VAO}
\end{equation}
and \(V_{\AO}(d)=0\) for \(d\ge2r\). The contact volume is \(V_{\AO}(0)=4\pi Rr^2+\frac{4\pi}{3}r^3\). The overlap volume decreases monotonically with increasing gap and is convex in \(d\) for \(R>r\). Near the outer edge \(d=2r\), \(V_{\AO}\sim \pi(R+r)(2r-d)^2\), so the volume vanishes quadratically and the force vanishes linearly. A deterministic trajectory started at \(d=2r-\varepsilon\) has an entry time that diverges logarithmically as \(\varepsilon\to0^+\); thermal fluctuations are therefore required to enter the depletion layer from \(d=2r\). These geometric facts are used throughout the paper.

The main results are as follows. (i) An exact AO sphere-plane excluded-volume geometry is derived with a thermodynamic interpretation. (ii) A thermodynamically consistent Smoluchowski--Robin (SR) formulation with variable diffusivity is obtained. (iii) Exact quadrature formulas are derived for the reflecting-outer mean first-passage time, the reaction-before-escape probability, and the conditional success time. (iv) The conditional success process is described by a Doob transform, an \(h\)-transform with \(h=u\) that makes explicit the selection bias of conditioning on successful capture. (v) Initial distributions and the selection bias implicit in conditioning on success are treated explicitly. (vi) Exact \(k_s\)-sensitivity formulas are derived, and \(d_c\)-sensitivity is analyzed separately. (vii) Independent numerical checks confirm the analytic results and the radiation-boundary mass balance.

In this work, we focus on the single-capture problem during the cell uptake. The AO geometry used here is the exact ideal hard-sphere geometry before contact. The CS replacement is a pressure-only approximation; the AO geometry itself is exact for hard-sphere crowders in the dilute limit.

\section{Geometry and thermodynamics}

\subsection{AO sphere-plane excluded-volume geometry}

Let the membrane be the plane \(z=0\), with the crowder-accessible side \(z>0\). A spherical particle of radius \(R\) has its lowest point at height \(d\ge0\) above the membrane. The particle center is at
\begin{equation}
z_C=R+d.
\label{eq:zC}
\end{equation}
A crowder of radius \(r\) cannot approach the particle surface closer than \(r\). The particle-induced exclusion region for crowder centers is therefore a sphere of radius
\begin{equation}
A=R+r
\label{eq:A}
\end{equation}
centered at \(z_C=R+d\). The wall-induced exclusion region for crowder centers is the half-space
\begin{equation}
z<r.
\label{eq:wallExclusion}
\end{equation}
When \(d<2r\), the intersection of the exclusion sphere and the wall-excluded half-space is a spherical cap.

The lowest point of the exclusion sphere is \(z_{\min}=z_C-A=d-r\), and the cap height is
\begin{equation}
q=r-z_{\min}=2r-d.
\label{eq:q}
\end{equation}
For \(d\ge2r\), the exclusion sphere does not overlap the wall-excluded half-space, and the AO overlap volume is zero. The geometric construction is shown in Fig.~\ref{fig:schematic}. Appendix~\hyperref[app:AO]{A} gives the explicit cap integration, the differentiation of \(V_{\AO}(d)\), and the outer-edge expansion.

\begin{figure*}[t]
  \centering
  \includegraphics[width=\linewidth]{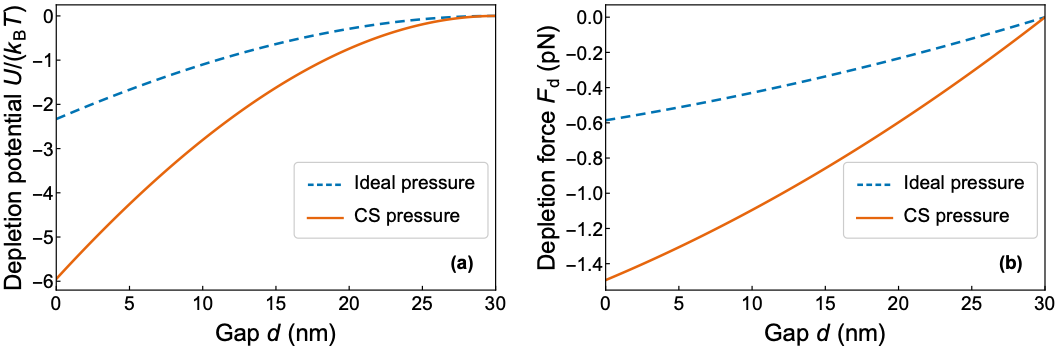}
  \caption{Depletion thermodynamics for the exact ideal hard-sphere AO geometry. (a) Depletion potential \(U/k_{\rm B}T\) versus gap \(d\) (nm) for the ideal van~'t~Hoff pressure and the CS pressure-only approximation. (b) Signed depletion force \(F_d\) along increasing gap \(d\) (nm) for the same two pressure models; negative values indicate attraction toward the membrane. In (a) and (b), both pressure choices vanish continuously at \(d=2r\), with \(V_{\AO}\propto(2r-d)^2\) and \(F_d\propto(2r-d)\) near the outer edge. Curves use \(R=50\,\mathrm{nm}\) and \(r=15\,\mathrm{nm}\).}
  \label{fig:ao_combined}
\end{figure*}

Integrating the horizontal disk areas of the spherical cap of radius \(A\) and height \(q\) gives \(V_{\rm cap}=\pi q^2(A-q/3)\). Substituting \(A=R+r\) and \(q=2r-d\) gives Eq.~\eqref{eq:VAO}. The contact volume follows at \(d=0\):
\begin{equation}
V_{\AO}(0)=4\pi Rr^2+\frac{4\pi}{3}r^3.
\label{eq:VAO0}
\end{equation}
Differentiating Eq.~\eqref{eq:VAO} with respect to \(d\) gives
\begin{equation}
V_{\AO}'(d)=-\pi(2r-d)(2R+d)\le0.
\label{eq:dVAO}
\end{equation}
The overlap volume decreases monotonically with increasing gap. Differentiating once more gives \(V_{\AO}''(d)=2\pi(R-r+d)>0\) for \(R>r\), which holds in the Table~\ref{tab:parameters} range; the overlap volume is convex in the physical gap. Near the outer edge \(d=2r\), writing \(s=2r-d\) gives
\begin{equation}
V_{\AO}(2r-s)=\pi(R+r)s^2-\frac{\pi}{3}s^3.
\label{eq:outerVAO}
\end{equation}
The volume vanishes quadratically at \(s=0\), and the force vanishes linearly. A deterministic trajectory started at \(d=2r-\varepsilon\) therefore has an entry time that diverges logarithmically as \(\varepsilon\to0^+\); thermal fluctuations are required to enter the depletion layer from \(d=2r\). The explicit cap integration, the differentiation, and the outer-edge expansion are collected in Appendix~\hyperref[app:AO]{A}.

\subsection{Depletion potential and force}

For an ideal crowder reservoir with osmotic pressure \(\Pi\), the AO free energy is
\begin{equation}
U(d)=-\Pi V_{\AO}(d),
\qquad U(d)=0\quad(d\ge2r).
\label{eq:U}
\end{equation}
The generalized force along increasing gap is
\begin{equation}
F_d(d)=-\frac{\dd U}{\dd d}=-\pi\Pi(2r-d)(2R+d)\le0.
\label{eq:F}
\end{equation}
The negative sign means that the force pushes the particle toward smaller \(d\), that is, toward the membrane. At contact,
\begin{equation}
|F_d(0)|=4\pi\Pi Rr.
\label{eq:F0}
\end{equation}
The potential is plotted in Fig.~\ref{fig:ao_combined}(a), and the force is plotted in Fig.~\ref{fig:ao_combined}(b). Both panels compare the ideal van~'t~Hoff pressure with the CS pressure-only approximation.

\subsection{Osmotic pressure models}

The AO potential \eqref{eq:U} is proportional to the osmotic pressure
\(\Pi\) of the crowder reservoir. For ideal crowders we take the van~'t~Hoff form
\begin{equation}
\Pi_{\rm id}=ck_{\rm B}T.
\label{eq:Pideal}
\end{equation}
This is the simplest ideal osmotic pressure expression, with the
osmotic coefficient set to unity. The same van~'t~Hoff expression has been used to describe the vestigial osmotic pressure produced by
internally generated daughter vesicles in a parent GUV
\cite{Wu2023Vestigial}; here it is used only as the ideal reference
pressure of the external crowder bath, while the CS
replacement below is a separate pressure-only approximation.

For a monodisperse hard-sphere fluid, the CS compressibility factor is
\begin{equation}
Z_{\CS}(\phi)
=\frac{1+\phi+\phi^2-\phi^3}{(1-\phi)^3},
\qquad
\phi=\frac{4\pi}{3}cr^3.
\label{eq:ZCS}
\end{equation}
The CS pressure is
\begin{equation}
\Pi_{\CS}=ck_{\rm B}T Z_{\CS}(\phi).
\label{eq:PCS}
\end{equation}
Replacing \(\Pi_{\rm id}\) by \(\Pi_{\CS}\) while keeping the ideal AO geometry is a pressure-only approximation:
\begin{equation}
U_{\CS}(d)=-\Pi_{\CS}V_{\AO}(d).
\label{eq:UCS}
\end{equation}
The pressure-only CS replacement is a neutral pressure correction: it changes the amplitude of the AO volume term but does not include wall layering or particle-correlation effects beyond the hard-sphere pressure. The AO geometry remains exact for hard-sphere crowders in the dilute limit.

For the values listed in Table~\ref{tab:parameters},
\begin{equation}
R_0=50\,\mathrm{nm},\
r_0=15\,\mathrm{nm},\
c_0=1.5\times10^{22}\,\mathrm{m^{-3}},\
T_0=300\,\mathrm{K}.
\label{eq:baseline}
\end{equation}
The corresponding volume fraction and compressibility factor are
\begin{equation}
\phi_0=\frac{4\pi}{3}c_0r_0^3=0.2120575,
\
Z_{\CS}(\phi_0)=2.5500782.
\label{eq:phiZ}
\end{equation}
The two pressure values are
\begin{equation}
\Pi_{\rm id}=62.1292\,\mathrm{Pa},
\qquad
\Pi_{\CS}=158.434\,\mathrm{Pa}.
\label{eq:pressures}
\end{equation}
The contact depth is
\begin{equation}
\beta|U_{\AO}(0)|=
\begin{cases}
2.3326, & \text{ideal pressure},\\
5.9484, & \text{CS pressure-only approximation}.
\end{cases}
\label{eq:depth}
\end{equation}
The contact volume is
\begin{equation}
V_{\AO}(0)=1.5550884\times10^{-22}\,\mathrm{m^3}.
\label{eq:V0}
\end{equation}

At fixed \(c\), increasing \(r\) changes the volume fraction as
\(\phi\propto r^3\). At \(r=20\,\mathrm{nm}\),
\begin{equation}
\phi=0.50265.
\label{eq:phi20}
\end{equation}
This value lies above the hard-sphere fluid freezing volume fraction
\(\phi_f\simeq0.494\) \cite{Pusey1986,Alder1957}, where the uniform-fluid
branch terminates. For \(\phi>\phi_f\), the CS
expression remains algebraically defined, but it no longer represents a
stable uniform fluid; it is used here only as a formal extrapolation of
the fluid branch. The endpoint \(r=20\,\mathrm{nm}\) is retained to
cover the Table~\ref{tab:parameters} range and is marked as a
fluid-branch extrapolation.

\section{Stochastic model}

\subsection{Markov assumptions and near-wall diffusivity}

The stochastic coordinate is the physical gap
\begin{equation}
d_t\in[a,b],
\qquad
a=d_c,\quad b=2r.
\label{eq:interval}
\end{equation}
The inner boundary \(d=a\) is the capture surface. The outer boundary \(d=b\) is the onset of the AO depletion layer. The model resolves this normal gap coordinate; lateral diffusion, orientation, membrane undulations, and discrete receptor positions are absorbed into an effective potential, an effective diffusivity, an effective reaction speed, and an initial distribution. The Markov assumption requires that crowder and solvent degrees of freedom relax faster than the gap coordinate and that memory effects can be neglected on the time scale of capture.

Near the membrane, the normal motion of the particle requires squeezing
out the fluid between the particle and the wall, so the normal
diffusivity decreases with the gap. For the calculations reported here,
we use the interpolation
\begin{equation}
D(d)=D_0\frac{d}{R+d},
\qquad
D_0=\frac{k_{\rm B}T}{6\pi\mu R},
\label{eq:Pade}
\end{equation}
where \(D_0\) is the bulk Stokes--Einstein (SE) diffusivity. This form reproduces the bulk limit \(D\to D_0\) for \(d\gg R\) and the lubrication scaling \(D\sim D_0d/R\) for \(d\ll R\).

\subsection{Free-energy functional and Smoluchowski equation}

Let \(p(d,t)\) be the probability density on \([a,b]\). The probability free energy is
\begin{equation}
\mathcal F[p]
=\int_a^b\left[U(d)p(d)+k_{\rm B}Tp(d)\ln p(d)\right]\dd d,
\label{eq:Ffree}
\end{equation}
up to an additive constant. Minimizing \(\mathcal F\) under the normalization \(\int_a^b p\,\dd d=1\) gives the Boltzmann weight \(p_{\rm eq}(d)=Z^{-1}\ee^{-\beta U(d)}\), where \(\beta=1/(k_{\rm B}T)\) and \(Z\) is the normalization constant. A small perturbation \(p\mapsto p+\epsilon\eta\) with \(\epsilon\ll1\) gives the variational chemical potential
\begin{equation}
\mu_p(d,t)
=\frac{\delta\mathcal F}{\delta p}
=U(d)+k_{\rm B}T[\ln p(d,t)+1].
\label{eq:mup}
\end{equation}
Let \(M(d)>0\) be the mobility and \(D(d)=k_{\rm B}TM(d)\) the diffusivity. The linear irreversible flux is \(J=-M(d)p\,\partial_d\mu_p\). Using \(\beta=1/(k_{\rm B}T)\),
\begin{equation}
J
=-M p\left[U'+k_{\rm B}T\frac{p'}{p}\right]
=-D\left[p'+\beta U'p\right].
\label{eq:J}
\end{equation}
Probability conservation \(\partial_t p=-\partial_d J\) gives the conserved Smoluchowski equation
\begin{equation}
\partial_t p
=\partial_d\left[D(d)\left(\partial_d p+\beta U'(d)p\right)\right].
\label{eq:Smol}
\end{equation}
At zero flux, \(p'+\beta U'p=0\), so the equilibrium density is
\begin{equation}
p_{\rm eq}(d)=Z^{-1}\ee^{-\beta U(d)}.
\label{eq:peq}
\end{equation}
The spatially varying diffusivity does not enter the equilibrium Boltzmann weight.

\subsection{It\^o equation and backward generator}

Expanding Eq.~\eqref{eq:Smol} gives the Fokker--Planck form
\begin{equation}
\partial_t p
=-\partial_d\left[(D'-\beta DU')p\right]+\partial_d^2(Dp).
\label{eq:FPexpanded}
\end{equation}
Comparing with the standard It\^o form
\begin{equation}
\partial_t p=-\partial_d(b p)+\partial_d^2(Dp)
\label{eq:FPstandard}
\end{equation}
gives the drift
\begin{equation}
b(d)=D'(d)-\beta D(d)U'(d).
\label{eq:ItoDrift}
\end{equation}
The equivalent It\^o stochastic differential equation is
\begin{equation}
\dd d_t=\left[D'(d_t)-\beta D(d_t)U'(d_t)\right]\dd t+\sqrt{2D(d_t)}\,\dd W_t.
\label{eq:Ito}
\end{equation}
The drift contains the noise-induced term \(D'\). If \(D'\) is omitted, the zero-flux equilibrium becomes proportional to \(D^{-1}\ee^{-\beta U}\), which is inconsistent with Eq.~\eqref{eq:peq}. For a smooth test function \(f\), the backward generator is
\begin{equation}
\mathcal L f=D f''+\left(D'-\beta DU'\right)f',
\label{eq:L}
\end{equation}
which can also be written in the self-adjoint form
\begin{equation}
\mathcal L f=\ee^{\beta U}\left(D\ee^{-\beta U}f'\right)'.
\label{eq:Lself}
\end{equation}

\begin{figure*}[t]
  \centering
  \includegraphics[width=\linewidth]{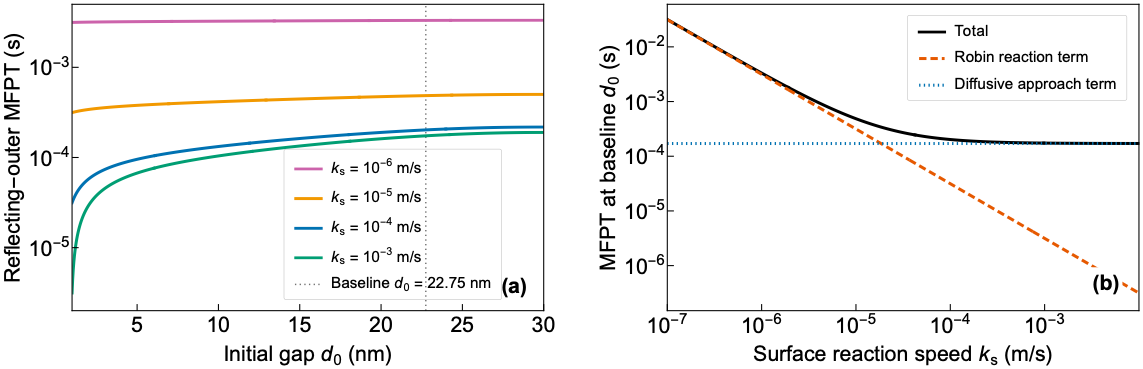}
  \caption{Reflecting-outer mean first-passage time and its reaction/diffusion decomposition. (a) \(\tau(d_0)\) versus initial gap \(d_0\) for several surface reaction speeds \(k_s\); the vertical dotted line marks the baseline \(d_0=22.75\,\mathrm{nm}\). Curves increase monotonically with \(d_0\) and have zero slope at the reflecting outer boundary \(b=2r\). (b) Total mean time, Robin reaction term, and diffusive approach term versus \(k_s\) at the baseline point. The reaction term scales as \(k_s^{-1}\), whereas the diffusion term is independent of \(k_s\); the sum is \(\tau(d)=A/k_s+B(d)\) of Eq.~\eqref{eq:tauAB}. Both panels use the same potential, near-wall diffusivity, and Table~\ref{tab:baseline_values} baseline values.}
  \label{fig:mfpt_combined}
\end{figure*}

\subsection{Robin boundary and backward conditions}

The inner boundary is a partially absorbing reaction surface. With \(J\) defined along increasing \(d\), the probability flux into the reaction surface is
\begin{equation}
J(a,t)=-k_s p(a,t),
\label{eq:J Robin}
\end{equation}
where \(k_s>0\) has dimensions of length per time. Equivalently,
\begin{equation}
D(a)\left[p'(a,t)+\beta U'(a)p(a,t)\right]=k_s p(a,t).
\label{eq:forward Robin}
\end{equation}
The outer boundary is either reflecting,
\begin{equation}
J(b,t)=0,
\label{eq:reflecting}
\end{equation}
or absorbing,
\begin{equation}
p(b,t)=0.
\label{eq:absorbing}
\end{equation}
The backward problem is obtained by integrating \(f\mathcal L^*p\) by parts, where \(\mathcal L^*p=\partial_d[D(p'+\beta U'p)]\) is the forward operator. The boundary terms vanish if
\begin{equation}
D(a)f'(a)=k_s f(a).
\label{eq:backRobin}
\end{equation}
This is the backward Robin condition for the mean first-passage time and for the homogeneous part of the conditional-success-time equation. For the reaction probability, the terminal reward is one at reaction, so the backward condition becomes
\begin{equation}
D(a)u'(a)=k_s[u(a)-1].
\label{eq:uRobin}
\end{equation}
At the reflecting outer boundary, \(f'(b)=0\); at the absorbing outer boundary, \(f(b)=0\). The full Green identity and the boundary-term elimination are given in Appendix~\hyperref[app:green]{B}.

\subsection{Dimensionless form}

Let
\begin{equation}
L=b-a=2r-d_c,
\qquad
x=\frac{d-a}{L}\in[0,1],
\label{eq:x}
\end{equation}
and choose a reference diffusivity \(D_{\rm ref}\), for example \(D_0=k_{\rm B}T/(6\pi\mu R)\). Define
\begin{equation}
\theta=\frac{D_{\rm ref}}{L^2}t,
\
q(x)=\frac{D(a+Lx)}{D_{\rm ref}},
\
W(x)=\beta U(a+Lx).
\label{eq:dimless}
\end{equation}
The dimensionless Smoluchowski equation is
\begin{equation}
\partial_\theta \rho
=\partial_x\left\{q(x)\left[\rho_x+W_x\rho\right]\right\}.
\label{eq:dimSmol}
\end{equation}
The inner flux condition becomes
\begin{equation}
j(0)=-\kappa_s\rho(0),
\qquad
\kappa_s=\frac{k_sL}{D_{\rm ref}}.
\label{eq:Da}
\end{equation}
The dimensionless backward operator is
\begin{equation}
\widehat{\mathcal L}f
=q f_{xx}+(q_x-qW_x)f_x
=\ee^{W}\partial_x\left(q\ee^{-W}f_x\right).
\label{eq:dimL}
\end{equation}
The surface Damk\"ohler number \(\kappa_s\) compares the surface reaction speed with cross-gap diffusion. A local Damk\"ohler number is
\begin{equation}
\kappa_a=\frac{k_sL}{D(a)}=\frac{\kappa_s}{q(0)}.
\label{eq:localDa}
\end{equation}
Because \(D(a)\) can be much smaller than \(D_0\) near the wall, \(\kappa_a\) can be large even when \(\kappa_s\) is moderate. The local Damk\"ohler number measures the boundary reaction rate relative to near-wall diffusion. It is not by itself the mean-time crossover parameter: from Eq.~\eqref{eq:tauAB}, the crossover satisfies \(k_s^{\rm cross}(d)=A/B(d)\), which depends on the whole potential, the diffusivity profile, and the initial position.

\section{First-passage quantities}

\subsection{Reflecting-outer mean first-passage time}

Let \(T_R\) be the first time at which the particle reacts at \(a\), with the outer boundary \(b\) reflecting. Define
\begin{equation}
\tau(d)=\E_d[T_R].
\label{eq:tau}
\end{equation}
The backward equation is
\begin{equation}
\mathcal L\tau=-1,
\qquad
a<d<b,
\label{eq:tauBVP}
\end{equation}
with
\begin{equation}
D(a)\tau'(a)=k_s\tau(a),
\qquad
\tau'(b)=0.
\label{eq:tauBC}
\end{equation}
Using the self-adjoint form \eqref{eq:Lself} and integrating twice gives
\begin{equation}
\begin{aligned}
\tau(d_0)
=&\frac{\ee^{\beta U(a)}}{k_s}\int_a^b\ee^{-\beta U(z)}\dd z \\
&+\int_a^{d_0}\frac{\ee^{\beta U(y)}}{D(y)}
\left[\int_y^b\ee^{-\beta U(z)}\dd z\right]\dd y.
\label{eq:tauDouble}
\end{aligned}
\end{equation}
The first term is independent of the initial point and scales as \(1/k_s\). The second term is the diffusion contribution. This decomposition is mathematical; it is not a decomposition into two independent physical trajectories. The derivation is given in Appendix~\hyperref[app:firstpassage]{C}. The mean time as a function of initial gap is shown in Fig.~\ref{fig:mfpt_combined}(a).

For \(a\le d_0<b\), the mean time is strictly increasing:
\begin{equation}
\tau'(d_0)
=\frac{\ee^{\beta U(d_0)}}{D(d_0)}
\int_{d_0}^b\ee^{-\beta U(z)}\dd z
>0.
\label{eq:tauMono}
\end{equation}
At the reflecting endpoint \(d_0=b\), \(\tau'(b)=0\). For fixed potential and diffusivity,
\begin{equation}
\frac{\partial\tau}{\partial k_s}
=-\frac{\ee^{\beta U(a)}}{k_s^2}
\int_a^b\ee^{-\beta U(z)}\dd z
<0.
\label{eq:tauKs}
\end{equation}
Thus the mean time is strictly decreasing and convex in \(k_s\). In the limit \(k_s\to\infty\),
\begin{equation}
\tau(d_0)\to
\int_a^{d_0}\frac{\ee^{\beta U(y)}}{D(y)}
\left[\int_y^b\ee^{-\beta U(z)}\dd z\right]\dd y,
\label{eq:tauAbsorb}
\end{equation}
the fully absorbing inner boundary result. In the limit \(k_s\to0^+\),
\begin{equation}
\tau(d_0)\sim
\frac{\ee^{\beta U(a)}}{k_s}
\int_a^b\ee^{-\beta U(z)}\dd z,
\label{eq:tauSlow}
\end{equation}
which diverges as \(1/k_s\). For a flat potential \(U=0\) and constant \(D\),
\begin{equation}
\tau(d_0)
=\frac{b-a}{k_s}
+\frac{(b-a)^2-(b-d_0)^2}{2D}.
\label{eq:tauFlat}
\end{equation}
At \(d_0=b\), \(\tau(b)=L/k_s+L^2/(2D)\).

\subsection{Reaction-before-escape probability}

\begin{figure*}[t]
  \centering
  \includegraphics[width=\linewidth]{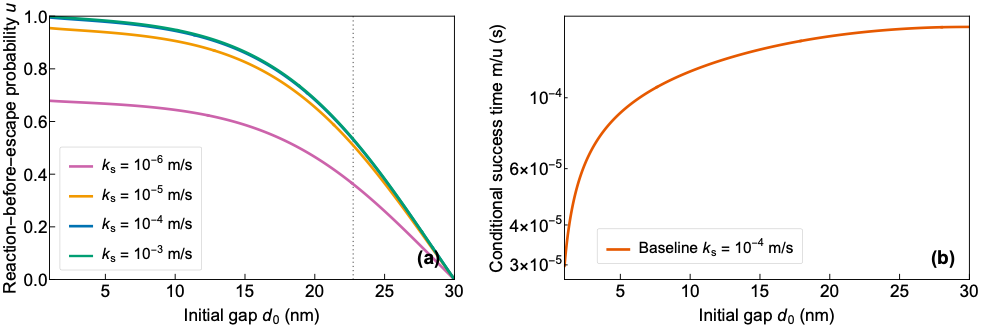}
  \caption{Reaction-before-escape probability and conditional success time with an absorbing outer boundary. (a) \(u(d_0)\) versus initial gap \(d_0\) for several surface reaction speeds \(k_s\); all curves satisfy \(u(b)=0\) and decrease with increasing \(d_0\), reflecting the competition between Robin reaction at \(a\) and escape at \(b\). (b) Conditional success time \(m/u\) versus \(d_0\) for the baseline \(k_s=10^{-4}\,\mathrm{m\,s^{-1}}\); the exact outer endpoint is not plotted because \(u=m=0\) there and the ratio is defined only by a one-sided limit. The conditional time is shorter than the reflecting-outer mean because only reaction-before-escape trajectories are averaged. Vertical axes are dimensionless in (a); the vertical axis in (b) is in seconds.}
  \label{fig:stochastic_combined}
\end{figure*}

Let \(b\) be an absorbing escape boundary. Define
\begin{equation}
u(d)=\Prb_d(T_R<T_E),
\label{eq:u}
\end{equation}
where \(T_R\) is the reaction time at \(a\) and \(T_E\) is the first hitting time of \(b\). The backward equation is
\begin{equation}
\mathcal L u=0,
\qquad
a<d<b,
\label{eq:uBVP}
\end{equation}
with
\begin{equation}
D(a)u'(a)=k_s[u(a)-1],
\qquad
u(b)=0.
\label{eq:uBC}
\end{equation}
Integrating the self-adjoint equation gives
\begin{equation}
u(d)
=\frac{k_s}{k_s+B}
\left[1-\frac{I(d)}{I(b)}\right],
\label{eq:uFinal}
\end{equation}
where \(I(x)=\int_a^x\ee^{\beta U(y)}/D(y)\dd y\) and \(B=\ee^{\beta U(a)}/I(b)\). The first factor describes the competition between reaction and return at the inner boundary. The second factor is the potential-weighted probability of reaching the inner boundary before escaping at \(b\), in the fully absorbing inner limit. The product form is exact for the one-dimensional stationary problem with a local Robin boundary. The derivation is given in Appendix~\hyperref[app:firstpassage]{C}. The reaction probability as a function of initial gap is shown in Fig.~\ref{fig:stochastic_combined}(a).

For a flat potential and constant \(D\), \(I(d)=(d-a)/D\) and \(B=D/(b-a)\), so
\begin{equation}
u(d)
=\frac{k_s(b-d)}{D+k_sL}.
\label{eq:uFlat}
\end{equation}
As \(k_s\to\infty\), \(u(d)\to(b-d)/L\), the classical splitting probability. As \(k_s\to0\), \(u(d)\to0\).

\subsection{Conditional success time}

Define the joint Laplace transform
\begin{equation}
G_s(d)
=\E_d\left[\ee^{-sT_R}\mathbf 1_{\{T_R<T_E\}}\right],
\qquad s\ge0.
\label{eq:Gs}
\end{equation}
Then \(G_0=u\). The backward equation is
\begin{equation}
(\mathcal L-s)G_s=0,
\label{eq:GsBVP}
\end{equation}
with
\begin{equation}
G_s(b)=0,
\qquad
D(a)G_s'(a)=k_s[G_s(a)-1].
\label{eq:GsBC}
\end{equation}
Define the unnormalized first moment
\begin{equation}
m(d)
=-\left.\frac{\partial G_s(d)}{\partial s}\right|_{s=0}
=\E_d\left[T_R\mathbf 1_{\{T_R<T_E\}}\right].
\label{eq:m}
\end{equation}
Differentiating Eq.~\eqref{eq:GsBVP} with respect to \(s\) and setting \(s=0\) gives
\begin{equation}
\mathcal L m=-u,
\label{eq:mBVP}
\end{equation}
with
\begin{equation}
m(b)=0,
\qquad
D(a)m'(a)=k_s m(a).
\label{eq:mBC}
\end{equation}
The conditional success time is
\begin{equation}
t_{\rm suc}(d)
=\E_d[T_R\mid T_R<T_E]
=\frac{m(d)}{u(d)},
\  u(d)>0.
\label{eq:tsuc}
\end{equation}
At \(b\), \(u=m=0\); the ratio is defined only by a limit. The resulting \(t_{\rm suc}(d_0)\) is plotted in Fig.~\ref{fig:stochastic_combined}(b). A cumulative-integral construction of \(m\) that avoids numerical cancellation is
\begin{equation}
m(a)=\frac{K}{1+k_s\ee^{-\beta U(a)}I_w},
\label{eq:ma}
\end{equation}
with \(I_w=\int_a^b w(y)\dd y\), \(K=\int_a^b w(y)\mathcal{A}(y)\dd y\), \(w(d)=\ee^{\beta U(d)}/D(d)\), and \(\mathcal{A}(d)=\int_a^d u(z)\ee^{-\beta U(z)}\dd z\). The derivation is given in Appendix~\hyperref[app:firstpassage]{C}.

\begin{figure*}[t]
  \centering
  \includegraphics[width=\linewidth]{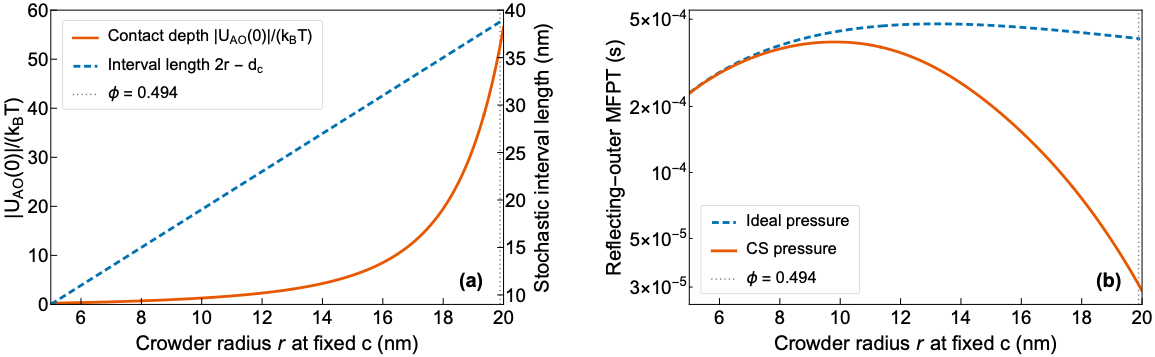}
  \caption{Fixed-\(c\) radius sweep. (a) CS pressure-only contact depth \(\beta|U_{\AO}(0)|\) (left vertical axis) and stochastic interval length \(2r-d_c\) (right vertical axis) versus crowder radius \(r\). (b) Reflecting-outer mean first-passage time versus \(r\) for the ideal and CS pressure-only approximations. The vertical dotted line in both panels marks the crowder radius at which \(\phi=0.494\). In this fixed-\(c\) sweep, pressure, geometry, interval length, initial position, and sampled mobility range change together, so the curves are not a pure pressure scan.}
  \label{fig:fixed_c_combined}
\end{figure*}

The conditional process is described by a Doob transform. For \(u>0\), define
\begin{equation}
\mathcal L^{(u)}f=\frac{1}{u}\mathcal L(uf).
\label{eq:Doob}
\end{equation}
Since \(\mathcal L u=0\), expanding the derivatives gives
\begin{equation}
\mathcal L^{(u)}f
=Df''+\left(D'-\beta DU'+2D\frac{u'}{u}\right)f'.
\label{eq:DoobL}
\end{equation}
The additional drift \(2Du'/u<0\) pushes the conditioned process away from the escape boundary and toward the reaction boundary. The conditional mean \(t_{\rm suc}=m/u\) satisfies
\begin{equation}
\mathcal L^{(u)}t_{\rm suc}=-1.
\label{eq:tsucDoob}
\end{equation}
The transformed inner boundary condition follows from \(m=ut_{\rm suc}\), \(D(a)m'(a)=k_sm(a)\), and \(D(a)u'(a)=k_s[u(a)-1]\):
\begin{equation}
D(a)t_{\rm suc}'(a)=\frac{k_s}{u(a)}t_{\rm suc}(a).
\label{eq:tsucBC}
\end{equation}
The outer endpoint is understood through the finite limit of \(m/u\); it is not a Dirichlet condition \(t_{\rm suc}(b)=0\). Near the escape boundary, \(u(d)\sim-u'(b)(b-d)\), so the Doob drift diverges as \(-2D(b)/(b-d)\). This is a selection effect of conditioning on success, not an additional physical force.

Note that the reflecting-outer mean time \(\tau\), the reaction-before-escape probability \(u\), and the conditional moment \(m\) satisfy different outer boundary conditions and should not be interchanged.

\subsection{Initial distributions and averaging}

The point initial condition \(d_0\) is a special case. If the initial gap has a normalized density \(\rho_0(d)\),
\begin{equation}
\int_a^b\rho_0(d)\dd d=1,
\label{eq:rho0}
\end{equation}
then the averaged reaction probability and mean first-passage time are
\begin{equation}
\bar u=\int_a^b u(d)\rho_0(d)\dd d,
\quad
\bar\tau=\int_a^b \tau(d)\rho_0(d)\dd d.
\label{eq:averages}
\end{equation}
The conditional success time averaged over the initial distribution is not \(\int \rho_0 m/u\). The correct expression is
\begin{equation}
\E_{\rho_0}[T_R\mid T_R<T_E]
=\frac{\int_a^b\rho_0(d)m(d)\dd d}
{\int_a^b\rho_0(d)u(d)\dd d}.
\label{eq:condAvg}
\end{equation}
The successful initial density is, by Bayes' rule,
\begin{equation}
\rho_0^{\rm succ}(d)
=\frac{\rho_0(d)u(d)}
{\int_a^b\rho_0(z)u(z)\dd z}.
\label{eq:rhosucc}
\end{equation}
Conditioning on success therefore reweights the initial distribution toward the reaction boundary.

Because \(\tau'(d)>0\) and \(u'(d)<0\), a distribution that is stochastically more outward has larger average mean time and smaller average reaction probability. The restricted equilibrium density
\begin{equation}
\rho_{\rm eq}(d)
=\frac{\ee^{-\beta U(d)}}
{\int_a^b\ee^{-\beta U(z)}\dd z}
\label{eq:rhoeq}
\end{equation}
is a possible preparation protocol. It is not automatically the initial distribution of the capture problem. It describes a fully relaxed interval with reflecting ends, not the reacting interval.

\section{Sensitivity to conditional parameters}

\begin{figure*}[t]
  \centering
  \includegraphics[width=\linewidth]{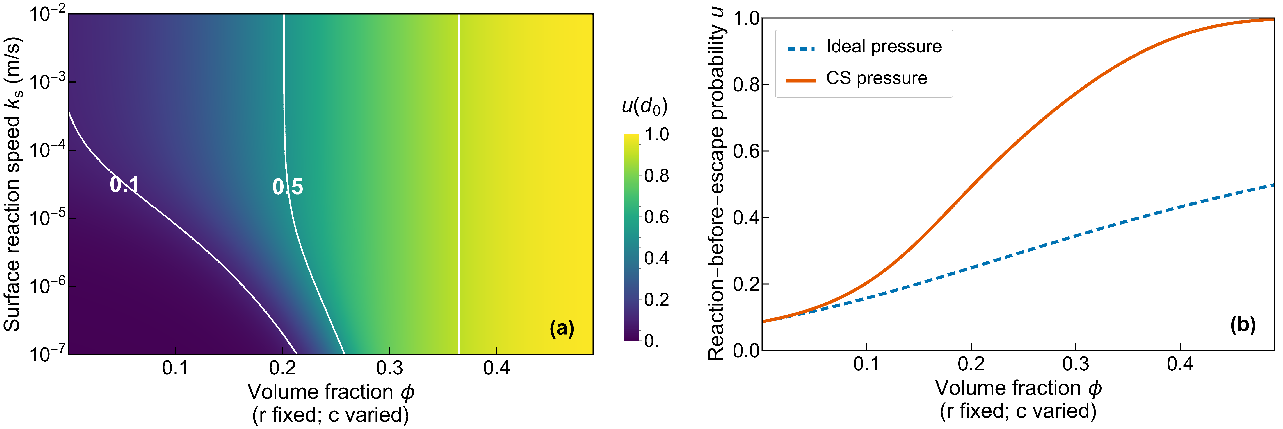}
  \caption{Conditional capture across crowder volume fraction and surface reaction speed at fixed geometry \(r=15\,\mathrm{nm}\). (a) Color map of \(u(d_0)\) at baseline \(d_0=22.75\,\mathrm{nm}\) in the \((\phi,k_s)\) plane, with \(\phi\) varied by changing \(c\) at fixed \(r\); white contours mark \(u=0.1,0.5,0.9\), and the probability increases toward larger \(\phi\) and \(k_s\). (b) \(u(d_0)\) versus \(\phi\) at fixed \(r\) for the ideal and CS pressure-only approximations at baseline \(k_s=10^{-4}\,\mathrm{m\,s^{-1}}\) and \(d_0=22.75\,\mathrm{nm}\); both curves increase with \(\phi\), and the CS closure lies above the ideal one because of the stronger depletion attraction. The scan varies \(c\) at fixed \(r\) and uses the same near-wall mobility closure; it isolates the pressure effect within this fixed-mobility model.}
  \label{fig:phi_combined}
\end{figure*}

\subsection{Capture distance \texorpdfstring{\(d_c\)}{dc}}

The capture distance \(d_c=a\) is not a Table~\ref{tab:parameters} parameter. Changing \(d_c\) changes the interval length \(L=2r-d_c\), the boundary potential \(U(a)\), the boundary diffusivity \(D(a)\), the relative initial position if \(d_0\) is held fixed, and the local Damk\"ohler number \(\kappa_a=k_sL/D(a)\). A \(d_c\)-scan is therefore not a scan of a single reaction-surface position. If the initial point is chosen by a fixed relative position,
\begin{equation}
d_0=a+\xi_0L,
\qquad
\xi_0\in(0,1),
\label{eq:d0rel}
\end{equation}
then
\begin{equation}
\frac{\partial d_0}{\partial d_c}=1-\xi_0.
\label{eq:dd0dc}
\end{equation}
If instead \(d_0\) is held fixed, the dimensionless initial position changes. The two protocols answer different questions. The sensitivity to \(d_c\) is multi-factor and is discussed here rather than displayed as a separate curve.

The Table~\ref{tab:parameters} complex size \(\delta=5\,\mathrm{nm}\) is not automatically equal to \(d_c\). A reaction boundary may depend on ligand extension, receptor height, orientational tolerance, membrane roughness, and the solvent layer. In this paper, \(d_c=1\,\mathrm{nm}\) is used only as a conditional choice, and \(d_c\)-sensitivity is reported separately.

\subsection{Surface reaction speed \texorpdfstring{\(k_s\)}{ks}}

The mean-time decomposition \eqref{eq:tauDouble} is written as
\begin{equation}
\tau(d)=\frac{A}{k_s}+B(d),
\label{eq:tauAB}
\end{equation}
where
\begin{equation}
A=\ee^{\beta U(a)}\int_a^b\ee^{-\beta U(z)}\dd z>0,
\label{eq:Adef}
\end{equation}
and
\begin{equation}
B(d)=\int_a^d\frac{\ee^{\beta U(y)}}{D(y)}
\int_y^b\ee^{-\beta U(z)}\dd z\dd y>0.
\label{eq:Bdef}
\end{equation}
Therefore
\begin{equation}
\frac{\partial\tau}{\partial k_s}=-\frac{A}{k_s^2},
\quad
\frac{\partial^2\tau}{\partial k_s^2}=\frac{2A}{k_s^3}>0.
\label{eq:tauKsDeriv}
\end{equation}
The reaction-limited fraction is
\begin{equation}
f_R(d)=\frac{A/k_s}{A/k_s+B(d)}.
\label{eq:fR}
\end{equation}
Then
\begin{equation}
\frac{\partial\ln\tau}{\partial\ln k_s}=-f_R(d).
\label{eq:tauLogSens}
\end{equation}
When \(f_R\approx1\), the mean time is reaction-controlled. When \(f_R\approx0\), it is diffusion-controlled. The crossover between these regimes is \(k_s^{\rm cross}(d)=A/B(d)\), not the local Damk\"ohler number alone. The decomposition is shown in Fig.~\ref{fig:mfpt_combined}(b).

For the reaction probability,
\begin{equation}
u(d)=u_\infty(d)\frac{k_s}{k_s+k_{\rm ret}},
\quad
k_{\rm ret}=\frac{\ee^{\beta U(a)}}{I(b)},
\label{eq:ukret}
\end{equation}
where
\begin{equation}
u_\infty(d)=1-\frac{I(d)}{I(b)}.
\label{eq:uinf}
\end{equation}
Therefore
\begin{equation}
\frac{\partial\ln u}{\partial\ln k_s}
=\frac{k_{\rm ret}}{k_s+k_{\rm ret}}.
\label{eq:uLogSens}
\end{equation}
This sensitivity is not the same function as the mean-time sensitivity. The two problems use different outer boundary conditions.

\begin{table*}[t]
\centering
\small
\renewcommand{\arraystretch}{1.08}
\caption{Baseline values derived from the Table~\ref{tab:parameters} inputs and the conditional parameters used for the numerical validation. The conditional parameters \(d_c\), \(k_s\), and \(d_0\) are not Table~\ref{tab:parameters} inputs. The finite-volume and Chebyshev values are independent numerical checks of the analytic quadrature. The last column identifies derived quantities, model parameters, model predictions, model decompositions, and numerical checks.}
\label{tab:baseline_values}
\begin{tabular*}{\textwidth}{@{\extracolsep{\fill}}l l l l@{}}
\toprule
Quantity & Value & Unit & How obtained \\
\midrule
\(R_0\) & \(50\) & nm & Table~\ref{tab:parameters} input \\
\(r_0\) & \(15\) & nm & Table~\ref{tab:parameters} input \\
\(c_0\) & \(1.5\times10^{22}\) & \(\mathrm{m^{-3}}\) & Table~\ref{tab:parameters} input \\
\(T_0\) & \(300\) & K & Table~\ref{tab:parameters} input \\
\(\phi_0\) & \(0.2120575\) & 1 & Derived: \(\frac{4\pi}{3}c_0r_0^3\) \\
\(Z_{\CS}(\phi_0)\) & \(2.5500782\) & 1 & Derived: CS EOS \\
\(\Pi_{\rm id}\) & \(62.1292\) & Pa & Derived: \(c_0k_{\rm B}T_0\) \\
\(\Pi_{\CS}\) & \(158.434\) & Pa & Derived: \(\Pi_{\rm id}Z_{\CS}\) \\
\(D_0\) & \(4.3947423\times10^{-12}\) & \(\mathrm{m^2\,s^{-1}}\) & Derived: SE \\
\(V_{\AO}(0)\) & \(1.5550884\times10^{-22}\) & \(\mathrm{m^3}\) & Derived: AO geometry \\
\(\beta|U_{\AO}(0)|\), ideal & \(2.3326\) & \(k_{\rm B}T\) & Derived \\
\(\beta|U_{\AO}(0)|\), CS & \(5.9484\) & \(k_{\rm B}T\) & Derived \\
\(d_c\) & \(1\) & nm & Conditional model parameter \\
\(k_s\) & \(10^{-4}\) & \(\mathrm{m\,s^{-1}}\) & Conditional model parameter \\
\(d_0\) & \(22.75\) & nm & Conditional model parameter \\
\(\tau\) (analytic) & \(2.0224365\times10^{-4}\) & s & Model prediction \\
\(\tau_{\rm react}\) & \(3.1467611\times10^{-5}\) & s & Model decomposition \\
\(\tau_{\rm diff}\) & \(1.7077604\times10^{-4}\) & s & Model decomposition \\
\(\tau\) (finite volume) & \(2.01813\times10^{-4}\) & s & Numerical check \\
\(\tau\) (spectral) & \(2.0224363\times10^{-4}\) & s & Numerical check \\
\(u(d_0)\) & \(0.53099275\) & 1 & Model prediction \\
\(t_{\rm suc}(d_0)\) & \(1.6101461\times10^{-4}\) & s & Model prediction \\
\bottomrule
\end{tabular*}
\end{table*}

\subsection{Parameter paths and coupled variations}

At fixed \(c\), changing \(r\) is a different path. It changes the volume fraction as
\begin{equation}
\phi(r)=\frac{4\pi}{3}cr^3,
\label{eq:phiFixedC}
\end{equation}
the CS pressure as
\begin{equation}
\Pi_{\CS}(r)=ck_{\rm B}TZ_{\CS}[\phi(r)],
\label{eq:PiFixedC}
\end{equation}
the overlap geometry through
\begin{equation}
V_{\AO}(d;R,r)=\frac{\pi}{3}(2r-d)^2(3R+r+d),
\label{eq:Vr}
\end{equation}
and the interval length through
\begin{equation}
L(r)=2r-d_c.
\label{eq:Lr}
\end{equation}
The initial position and the sampled mobility range also change if \(d_0\) is tied to the interval. This fixed-\(c\) radius scan is therefore not a pure pressure scan. The contact-depth and interval-length changes are shown in Fig.~\ref{fig:fixed_c_combined}(a), and the corresponding reflecting-outer mean times are shown in Fig.~\ref{fig:fixed_c_combined}(b).

At fixed crowder radius \(r\), changing the crowder concentration \(c\) changes the volume fraction and the osmotic pressure while leaving the geometry \(V_{\AO}(d;R,r)\), the interval length \(L=2r-d_c\), and the mobility profile \(D(d)\) unchanged. The volume fraction is
\begin{equation}
\phi(c,r)=\frac{4\pi}{3}cr^3,
\label{eq:phiFixedR}
\end{equation}
and the CS pressure is
\begin{equation}
\Pi_{\CS}(c,r)=ck_{\rm B}TZ_{\CS}[\phi(c,r)].
\label{eq:PiFixedR}
\end{equation}
This is the fixed-geometry crowding path. The conditional capture probability on this path is shown in Fig.~\ref{fig:phi_combined}(a) in the \((\phi,k_s)\) plane, and the ideal versus CS comparison at fixed \(r\) is shown in Fig.~\ref{fig:phi_combined}(b). Both panels vary \(c\) at fixed \(r\); they do not vary \(r\) at fixed \(\phi\).

A third distinct path is to fix \(\phi\) and change \(r\). This requires
\begin{equation}
c(\phi,r)=\frac{3\phi}{4\pi r^3}.
\label{eq:cphi}
\end{equation}
The pressure then scales as \(r^{-3}\) at fixed \(\phi\), while the depletion range scales as \(r\). This fixed-\(\phi\) radius path is discussed here for completeness but is not the path shown in Fig.~\ref{fig:phi_combined}; that figure varies \(c\) at fixed \(r\). The fixed-\(c\) radius scan of the mean time is shown in Fig.~\ref{fig:fixed_c_combined}(b).

\section{Numerical verification}

The analytic quadrature results in this paper were checked against independent finite-volume and spectral solutions of the same boundary-value problem. The survival probability, reaction flux, mean first-passage time, reaction probability, and conditional success time agreed within numerical discretization errors, and the discrete mass balance at the radiation boundary was satisfied. Since the focus here is on the physical picture and the analytical results, the algorithmic details of these checks are omitted. The baseline conditional parameters are \(d_c=1\,\mathrm{nm}\), \(k_s=10^{-4}\,\mathrm{m\,s^{-1}}\), and \(d_0=22.75\,\mathrm{nm}\). The analytic quadrature gives
\begin{equation}
\begin{aligned}
\tau&=2.0224365\times10^{-4}\,\mathrm{s},
\\
\tau_{\rm react}&=3.1467611\times10^{-5}\,\mathrm{s},
\\
\tau_{\rm diff}&=1.7077604\times10^{-4}\,\mathrm{s},
\label{eq:baselineTau}
\end{aligned}
\end{equation}
and
\begin{equation}
\begin{aligned}
u(d_0)&=0.53099275,
\\
t_{\rm suc}(d_0)&=1.6101461\times10^{-4}\,\mathrm{s}.
\label{eq:baselineU}
\end{aligned}
\end{equation}
The independent numerical checks gave values consistent with these results (Table~\ref{tab:baseline_values}). The parameters from Table~\ref{tab:parameters} and the conditional parameters used here are collected in Table~\ref{tab:baseline_values}.

\section{Results and discussion}

The values reported here illustrate the competition between reaction and diffusion. The reaction term \(\tau_{\rm react}=3.1467611\times10^{-5}\,\mathrm{s}\) is smaller than the diffusion term \(\tau_{\rm diff}=1.7077604\times10^{-4}\,\mathrm{s}\). The reflecting-outer mean time is therefore diffusion-dominated for the parameter values used here. The reaction-before-escape probability \(u(d_0)=0.53099275\) is close to one half: although the initial point is geometrically closer to the escape boundary than to the reaction boundary, the depletion attraction toward the membrane raises the capture probability relative to the pure-diffusion splitting value. With an absorbing outer boundary, the conditional success time \(1.6101461\times10^{-4}\,\mathrm{s}\) is shorter than the reflecting-outer mean \(2.0224365\times10^{-4}\,\mathrm{s}\). The two quantities use different outer boundary protocols: the former averages only reaction-before-escape trajectories, whereas the latter allows reflection at \(b\) and eventual reaction. The Doob transform makes the selection effect explicit; its additional drift \(2Du'/u\) pushes the conditioned process toward the reaction boundary.

The \(k_s\)-sensitivity is exact. From Eq.~\eqref{eq:tauAB}, the mean time is a sum of a \(1/k_s\) reaction term and a \(k_s\)-independent diffusion term. The reaction probability has the factor \(k_s/(k_s+k_{\rm ret})\) and saturates as \(k_s\to\infty\). The mean time and reaction probability therefore respond differently to \(k_s\). Their sensitivities are not inferred from one another.

The \(d_c\)-sensitivity is multi-factor. Reducing \(d_c\) shortens the interval, moves the reaction boundary into a deeper part of the AO well, changes the local diffusivity, and changes the relative initial position. These effects can partially cancel. A \(d_c\)-scan therefore specifies whether \(d_0\) is held fixed or is scaled with the interval. The Table~\ref{tab:parameters} complex size \(\delta=5\,\mathrm{nm}\) is not automatically \(d_c\). Identifying them would require an independent molecular-geometry argument. The parameter choice \(d_c=1\,\mathrm{nm}\) is therefore a conditional parameter, not a Table~\ref{tab:parameters} input.

The CS pressure-only approximation changes the depth of the AO well and does not change the geometry. With the parameter values adopted here, the CS contact depth is about \(5.95\,k_{\rm B}T\) instead of \(2.33\,k_{\rm B}T\). This increases the depletion attraction and shortens the diffusion part of the capture time. At \(r=20\,\mathrm{nm}\), \(\phi=0.50265\), which is beyond the hard-sphere freezing volume fraction. The CS curve is used there as a formal extrapolation of the fluid branch. The AO geometry itself is exact for hard-sphere crowders in the dilute limit; the pressure-only CS replacement is a pressure correction and should not be interpreted as a full finite-density depletion potential.

\section{Conclusion}

A thermodynamically consistent SR theory has been formulated for the pre-contact capture of a spherical particle by a planar membrane in a crowded environment. The geometry is the exact ideal hard-sphere AO sphere-plane excluded-volume geometry. The pressure is either ideal or CS pressure-only. The stochastic dynamics includes the noise-induced drift required by the spatially varying diffusivity. Exact quadrature formulas have been derived for the reflecting-outer mean first-passage time, the reaction-before-escape probability, and the conditional success time. The conditional process has been described by a Doob transform. The dependence on the surface reaction speed is explicit. The dependence on the capture distance is multi-factor. Independent numerical checks confirm the analytic results and the radiation-boundary mass balance.

The parameters in Table~\ref{tab:parameters} separate cleanly into thermodynamic inputs and conditional parameters. The AO contact volume, the ideal and CS osmotic pressures, the contact depths, and the bulk diffusivity are fixed by Table~\ref{tab:parameters}. The capture distance, the surface reaction speed, the initial gap, and the initial distribution are conditional. The mean time, the reaction probability, and the conditional success time are therefore reported as functions of these conditional parameters, not as unique predictions. This separation allows the theory to be compared with experiments or with more detailed simulations: a change in \(d_c\), \(k_s\), \(d_0\), or \(\rho_0\) can produce a different observable without any change in the Table~\ref{tab:parameters} physics.

A companion paper \cite{Wu2026multi} addresses the multivalent bond stabilization after the first bond and the coupling of the normal capture coordinate to lateral receptor diffusion and membrane deformation. The present theory provides a transparent and reproducible starting point for that extension.

\begin{acknowledgments}
H.W. thanks J. Liu and J. Xu for helpful discussions. This work is supported by the General Program of National Natural Science Foundation of China under Grant No. 12374210, the open research fund of Songshan Lake Materials Laboratory No. 2023SLABFN20, and the startup fund No. WIUCASQD2022005 from the Wenzhou Institute, University of Chinese Academy of Sciences.
\end{acknowledgments}

\appendix
\renewcommand{\thesection}{\Alph{section}}
\setcounter{section}{0}
\renewcommand{\theequation}{S\arabic{equation}}
\setcounter{equation}{0}
\renewcommand{\thefigure}{S\arabic{figure}}
\setcounter{figure}{0}

\section{Appendix A: AO excluded-volume geometry}
\label{app:AO}

The exclusion sphere has radius \(A=R+r\) and center at \(z_C=R+d\). The cutting plane is \(z=r\). The cap height is
\begin{equation}
q=r-(z_C-A)=2r-d.
\label{eq:appA:q}
\end{equation}
For a cap of height \(q\) in a sphere of radius \(A\),
\begin{equation}
V_{\rm cap}
=\int_0^q\pi(2As-s^2)\dd s
=\pi q^2\left(A-\frac{q}{3}\right).
\label{eq:appA:cap}
\end{equation}
Substituting \(A=R+r\) and \(q=2r-d\) gives Eq.~\eqref{eq:VAO}. The contact volume follows at \(d=0\):
\begin{equation}
V_{\AO}(0)=4\pi Rr^2+\frac{4\pi}{3}r^3.
\label{eq:appA:V0}
\end{equation}

To differentiate Eq.~\eqref{eq:VAO}, write \(u=2r-d\) and \(v=3R+r+d\). Then \(u'=-1\), \(v'=1\), and
\begin{equation}
V_{\AO}'(d)
=\frac{\pi}{3}\left(2uu'v+u^2v'\right)
=\frac{\pi}{3}u(u-2v).
\label{eq:appA:der1}
\end{equation}
Since
\begin{equation}
u-2v=(2r-d)-2(3R+r+d)=-3(2R+d),
\label{eq:appA:u2v}
\end{equation}
we obtain
\begin{equation}
V_{\AO}'(d)
=-\pi(2r-d)(2R+d)\le0.
\label{eq:appA:dV}
\end{equation}
Differentiating once more,
\begin{equation}
V_{\AO}''(d)
=-\pi\frac{\dd}{\dd d}\left[(2r-d)(2R+d)\right]
=2\pi(R-r+d).
\label{eq:appA:d2V}
\end{equation}
For \(R>r\), which holds in the Table~\ref{tab:parameters} range, \(V_{\AO}''>0\). The overlap volume is convex in the physical gap.

Near the outer edge \(d=2r\), write \(s=2r-d\). Then \(3R+r+d=3R+3r-s\), so
\begin{equation}
V_{\AO}(2r-s)
=\frac{\pi}{3}s^2(3R+3r-s)
=\pi(R+r)s^2-\frac{\pi}{3}s^3.
\label{eq:appA:outer}
\end{equation}
The volume vanishes quadratically at \(s=0\), and the force vanishes linearly.

\section{Appendix B: Robin boundary and Green identity}
\label{app:green}

The forward operator is
\begin{equation}
\mathcal L^*p=\partial_d\left[D(p'+\beta U'p)\right].
\label{eq:appB:Lstar}
\end{equation}
Let \(q=D(p'+\beta U'p)\). Integrating \(f\mathcal L^*p\) by parts,
\begin{equation}
\int_a^b f\mathcal L^*p\,\dd d
=\int_a^b f q'\,\dd d
=[fq]_a^b-\int_a^b f'q\,\dd d.
\label{eq:appB:Green1}
\end{equation}
Now
\begin{equation}
-\int_a^b f'q\,\dd d
=-\int_a^b f'D(p'+\beta U'p)\,\dd d.
\label{eq:appB:Green2}
\end{equation}
Integrating the \(f'Dp'\) term by parts,
\begin{equation}
-\int_a^b f'Dp'\,\dd d
=-[f'Dp]_a^b+\int_a^b p(Df')'\,\dd d.
\label{eq:appB:Green3}
\end{equation}
The remaining term is
\begin{equation}
-\int_a^b \beta DU'f'p\,\dd d.
\label{eq:appB:Green4}
\end{equation}
Combining,
\begin{equation}
\begin{aligned}
\int_a^b f\mathcal L^*p\,\dd d
=&[fD(p'+\beta U'p)-pDf']_a^b \\
&+\int_a^b p\left[(Df')'-\beta DU'f'\right]\dd d.
\label{eq:appB:Green5}
\end{aligned}
\end{equation}
The volume term is
\begin{equation}
(Df')'-\beta DU'f'
=Df''+(D'-\beta DU')f'
=\mathcal L f.
\label{eq:appB:GreenVolume}
\end{equation}
Therefore
\begin{equation}
\int_a^b f\mathcal L^*p\,\dd d
=[fD(p'+\beta U'p)-pDf']_a^b
+\int_a^b p\,\mathcal L f\,\dd d.
\label{eq:appB:Green}
\end{equation}
At the inner boundary, using \(J(a)=-k_s p(a)\),
\begin{equation}
\begin{aligned}
&f(a)D(a)(p'(a)+\beta U'(a)p(a))-p(a)D(a)f'(a) \\
=&p(a)\left[k_s f(a)-D(a)f'(a)\right].
\label{eq:appB:GreenInner}
\end{aligned}
\end{equation}
To remove the boundary term for arbitrary \(p(a)\), one imposes
\begin{equation}
D(a)f'(a)=k_s f(a).
\label{eq:appB:backRobin}
\end{equation}
This is the backward Robin condition for the mean first-passage time and for the homogeneous part of the conditional-success-time equation. For the reaction probability, the terminal reward is one at reaction, so the backward condition becomes
\begin{equation}
D(a)u'(a)=k_s[u(a)-1].
\label{eq:appB:uRobin}
\end{equation}
At the reflecting outer boundary, \(f'(b)=0\); at the absorbing outer boundary, \(f(b)=0\).

\section{Appendix C: First-passage quantities}
\label{app:firstpassage}

\subsection{C.1 Mean first-passage time}

Using the self-adjoint form \eqref{eq:Lself},
\begin{equation}
\ee^{\beta U}\left(D\ee^{-\beta U}\tau'\right)'=-1.
\label{eq:appC:tauFirst}
\end{equation}
Multiplying by \(\ee^{-\beta U}\),
\begin{equation}
\left(D\ee^{-\beta U}\tau'\right)'=-\ee^{-\beta U}.
\label{eq:appC:tauFirst2}
\end{equation}
Integrating from \(d\) to \(b\),
\begin{equation}
D(b)\ee^{-\beta U(b)}\tau'(b)-D(d)\ee^{-\beta U(d)}\tau'(d)
=-\int_d^b\ee^{-\beta U(z)}\dd z.
\label{eq:appC:tauIntegral1}
\end{equation}
The reflecting condition at \(b\) gives \(\tau'(b)=0\). Hence
\begin{equation}
D(d)\ee^{-\beta U(d)}\tau'(d)
=\int_d^b\ee^{-\beta U(z)}\dd z,
\label{eq:appC:tauPrime}
\end{equation}
and
\begin{equation}
\tau'(d)
=\frac{\ee^{\beta U(d)}}{D(d)}\int_d^b\ee^{-\beta U(z)}\dd z
\ge0.
\label{eq:appC:tauPrime2}
\end{equation}
Integrating from \(a\) to \(d_0\),
\begin{equation}
\tau(d_0)-\tau(a)
=\int_a^{d_0}\frac{\ee^{\beta U(y)}}{D(y)}
\left[\int_y^b\ee^{-\beta U(z)}\dd z\right]\dd y.
\label{eq:appC:tauIntegral2}
\end{equation}
At \(a\), using Eq.~\eqref{eq:appC:tauPrime2} in the Robin condition,
\begin{equation}
k_s\tau(a)
=D(a)\tau'(a)
=\ee^{\beta U(a)}\int_a^b\ee^{-\beta U(z)}\dd z.
\label{eq:appC:tauA1}
\end{equation}
Therefore
\begin{equation}
\tau(a)=\frac{\ee^{\beta U(a)}}{k_s}\int_a^b\ee^{-\beta U(z)}\dd z.
\label{eq:appC:tauA}
\end{equation}
Combining Eqs.~\eqref{eq:appC:tauIntegral2} and \eqref{eq:appC:tauA} gives Eq.~\eqref{eq:tauDouble}.

\subsection{C.2 Reaction-before-escape probability}

Using the self-adjoint form,
\begin{equation}
\left(D\ee^{-\beta U}u'\right)'=0.
\label{eq:appC:uFirst}
\end{equation}
Therefore
\begin{equation}
D(d)\ee^{-\beta U(d)}u'(d)=C,
\label{eq:appC:uC}
\end{equation}
where \(C\) is a constant. Define
\begin{equation}
I(x)=\int_a^x\frac{\ee^{\beta U(y)}}{D(y)}\dd y.
\label{eq:appC:I}
\end{equation}
Integrating Eq.~\eqref{eq:appC:uC},
\begin{equation}
u(d)=u(a)+CI(d).
\label{eq:appC:uI}
\end{equation}
The outer condition \(u(b)=0\) gives
\begin{equation}
C=-\frac{u(a)}{I(b)}.
\label{eq:appC:C}
\end{equation}
At the inner boundary,
\begin{equation}
D(a)u'(a)=C\ee^{\beta U(a)}.
\label{eq:appC:uPrimeA}
\end{equation}
Substituting into the Robin condition and defining
\begin{equation}
B=\frac{\ee^{\beta U(a)}}{I(b)},
\label{eq:appC:B}
\end{equation}
one obtains
\begin{equation}
-Bu(a)=k_s[u(a)-1],
\label{eq:appC:uaEquation}
\end{equation}
and therefore
\begin{equation}
u(a)=\frac{k_s}{k_s+B}.
\label{eq:appC:ua}
\end{equation}
Combining Eqs.~\eqref{eq:appC:uI}, \eqref{eq:appC:C}, and \eqref{eq:appC:ua} gives Eq.~\eqref{eq:uFinal}.

\subsection{C.3 Conditional moment and Doob transform}

Differentiating Eq.~\eqref{eq:GsBVP} with respect to \(s\),
\begin{equation}
\mathcal L(\partial_sG_s)-G_s-s\partial_sG_s=0.
\label{eq:appC:GsDerivative}
\end{equation}
Setting \(s=0\) and using \(G_0=u\) and \(m=-\partial_sG_s|_{s=0}\) gives Eq.~\eqref{eq:mBVP}. To construct \(m\) explicitly, write
\begin{equation}
\left(D\ee^{-\beta U}m'\right)'=-u\ee^{-\beta U}.
\label{eq:appC:mFirst}
\end{equation}
Define
\begin{equation}
\mathcal{A}(d)=\int_a^d u(z)\ee^{-\beta U(z)}\dd z,
\quad
w(d)=\frac{\ee^{\beta U(d)}}{D(d)}.
\label{eq:appC:Aw}
\end{equation}
Integrating Eq.~\eqref{eq:appC:mFirst} once,
\begin{equation}
D(d)\ee^{-\beta U(d)}m'(d)=C-\mathcal{A}(d),
\label{eq:appC:mFirstIntegral}
\end{equation}
where \(C\) is a constant. Therefore
\begin{equation}
m'(d)=w(d)[C-\mathcal{A}(d)].
\label{eq:appC:mPrime}
\end{equation}
The Robin condition gives
\begin{equation}
C=k_s\ee^{-\beta U(a)}m(a).
\label{eq:appC:Cm}
\end{equation}
Integrating Eq.~\eqref{eq:appC:mPrime} from \(a\) to \(b\) and using \(m(b)=0\),
\begin{equation}
0=m(a)+C\int_a^b w(y)\dd y-\int_a^b w(y)\mathcal{A}(y)\dd y.
\label{eq:appC:mIntegral}
\end{equation}
Let
\begin{equation}
I_w=\int_a^b w(y)\dd y,
\quad
K=\int_a^b w(y)\mathcal{A}(y)\dd y.
\label{eq:appC:IK}
\end{equation}
Substituting Eq.~\eqref{eq:appC:Cm} into Eq.~\eqref{eq:appC:mIntegral},
\begin{equation}
m(a)\left[1+k_s\ee^{-\beta U(a)}I_w\right]=K,
\label{eq:appC:maEquation}
\end{equation}
and hence Eq.~\eqref{eq:ma}.

For the Doob transform, expanding the derivatives in \(\mathcal L(uf)\) gives
\begin{equation}
(uf)'=u'f+uf',
\quad
(uf)''=u''f+2u'f'+uf''.
\label{eq:appC:DoobExpand2}
\end{equation}
Therefore
\begin{equation}
\begin{aligned}
\mathcal L(uf)
=&u\left[Df''+(D'-\beta DU')f'\right]
+2Du'f' \\
&+f\left[Du''+(D'-\beta DU')u'\right].
\label{eq:appC:DoobExpand3}
\end{aligned}
\end{equation}
The bracket in the last term is \(\mathcal L u=0\). Thus
\begin{equation}
\mathcal L(uf)=u\mathcal L f+2Du'f',
\label{eq:appC:DoobExpand4}
\end{equation}
and
\begin{equation}
\mathcal L^{(u)}f
=Df''+\left(D'-\beta DU'+2D\frac{u'}{u}\right)f'.
\label{eq:appC:DoobL}
\end{equation}
The additional drift \(2Du'/u<0\) pushes the conditioned process away from the escape boundary and toward the reaction boundary. The conditional mean \(t_{\rm suc}=m/u\) satisfies Eq.~\eqref{eq:tsucDoob} and the boundary condition Eq.~\eqref{eq:tsucBC}.

\end{document}